\documentclass[aps,pra,twocolumn,floatfix,showpacs,showkeys,nofootinbib]{revtex4-1}
\usepackage{graphicx}
\usepackage{natbib}
\usepackage{epstopdf}
\usepackage{amsmath}
\usepackage{amssymb}
\usepackage{mathbbol}
\usepackage{braket}
\usepackage{hyperref}

\usepackage{xcolor}

\begin{document}

%% title, authors etc.
\title{Probing the critical point of the Jaynes-Cummings second-order dissipative quantum phase transition}

\author{Th. K. Mavrogordatos}\email{t.mavrogordatos@ucl.ac.uk}
\affiliation{Department of Physics and Astronomy, University College London, Gower Street, London, WC1E 6BT, United Kingdom}

\date{\today}

\begin{abstract}
We highlight the importance of quantum fluctuations in organizing a dissipative quantum phase transition for the driven Jaynes-Cummings interaction with variable qubit-cavity detuning. The system response presents a substantial difference from the predictions of the semiclassical theory, the extent of which is revealed in the properties of quantum bistability, and visualized with the help of {\it quasi}-distribution functions for the cavity field, subject to an appropriate scale parameter. States anticipated by the neoclassical theory of radiation coexist in the quantum picture, following the occurrence of spontaneous dressed-state polarization and phase bistability at resonance.
\end{abstract}

\maketitle

The recent experimental demonstration of the breakdown of photon blockade \cite{FinkPhotonBlockade} by means of a dissipative quantum phase transition, following its theoretical prediction shortly beforehand \cite{CarmichaelPhotonBlockade}, marks an active interest in the study of phenomena where quantum fluctuations shape the system dynamics, yielding a response that cannot be interpreted as the superposition of a small amount of noise to the semiclassical predictions (see for example \cite{KimbleBlockade, Murch}). Under suitable conditions, bosons may exhibit a fermionic behaviour, with the photon blockade \textemdash{coined} as the analogue of the Coulomb blockade \cite{PhotonBlockadeImamoglu} \textemdash{and} the suppression of double occupancy in an array of coupled resonators being characteristic examples where a significant nonlinearity is responsible for the development of strong correlations in the spectrum of the system \cite{CarusottoPRL, CarmichaelPhotonBlockade}. The regime of photon blockade has been recently accessed to extract the fluorescence spectrum of a collection of coupled driven resonators in a dissipative environment with reference to an effective Ising chain model \cite{KildaFl, XYmodel, Majorana}.  

The forced Jaynes-Cummings (JC) oscillator, with one resonant and driven cavity mode coupled to a two-level system (qubit), exhibits a characteristic $\sqrt{n}$ nonlinearity \cite{JCoriginalpaper} dependent on the drive (see e.g. \cite{Bishop, CarmichaelPhotonBlockade}). When placed in a dissipative environment it demonstrates a fundamental out-of-equilibrium light-matter interaction where quantum phase transitions can be encountered \cite{singleatom, CarmichaelPhotonBlockade}. In this Letter, we explore the properties of the JC nonlinearity in the region of a second-order phase transition, varying the inter (system-environment) and intra (qubit-cavity, detuned) coupling strengths, having as a pivotal reference the critical point for spontaneous symmetry breaking \cite{SpontaneousDressed}. This point has a well-defined position in the parameter space of the drive at resonance, with its frequency equal to that of the qubit resonant with the cavity, and with its strength equal to the half of their coupling strength. It signals the collapse of the $\sqrt{n}$ nonlinearity to zero where the discrete {\it quasi}-energy spectrum of the JC oscillator merges to a continuum above threshold \cite{CarmichaelPhotonBlockade}. 

The Master Equation (ME) in the interaction picture and the rotating wave approximation (RWA) for a two-level atom with frequency $\omega_q$ and raising (lowering) operators $\sigma_{+} (\sigma_{-})$ interacting with a single cavity mode with frequency $\omega_c$ and raising (lowering) operators $a^{\dagger} (a)$, driven by a coherent field with strength $\varepsilon_d$ and frequency $\omega_d$ in the presence of dissipation at zero temperature, reads:
\begin{equation}\label{ME}
\begin{aligned}
&\dot{\rho}=i\Delta\omega_c[a^{\dagger}a, \rho] + i\Delta\omega_q[\sigma_{+}\sigma_{-}, \rho] - i \varepsilon_d[a+a^{\dagger}, \rho]\\ 
&-ig[a\sigma_{+}+a^{\dagger}\sigma_{-}, \rho] + \kappa \left(2a\rho a^{\dagger}-a^{\dagger}a \rho - \rho a^{\dagger}a\right) \\
&+ (\gamma/2) \left(2\sigma_{-}\rho \sigma_{+} - \sigma_{+}\sigma_{-}\rho - \rho \sigma_{+} \sigma_{-}\right),
\end{aligned}
\end{equation}
where $\Delta \omega_{(c,q)}=\omega_d-\omega_{(c,q)}$ is the detuning between the laser driving field and the (cavity field, qubit) respectively. The coupling strength between the cavity mode and the atom, detuned by $\delta \equiv \omega_q-\omega_c$ [with $|\delta| \ll \omega_{(c,q)}$], is denoted by $g$ [with $g \ll \omega_{(c,q)}$], an interaction which is assumed to be much stronger than the cavity decay rate $2\kappa$ and the spontaneous emission rate $\gamma$ in the {\it strong coupling regime}. The ME \eqref{ME} does not yield a closed set of equations for the first-order moments of the field and the two-level atom. Factorizing higher (than first) order moments results in neglecting quantum correlations, and produces a closed set of equations called the {\it Maxwell-Bloch} equations \cite{WallsBook}. 

We now require a means to depict the quantum fluctuations for the density matrices satisfying Eq. \ref{ME}. The $Q$ function in the steady state: 
\begin{equation}\label{Qfunc}
Q(x+iy)=\frac{1}{\pi}\braket{x+iy|\rho_{cv, {\rm ss}}|x+iy}
\end{equation}
is used to provide a ``classical'' visualization of the intracavity radiation field in the quantum-classical correspondence provided by {\it quasi}-probability distributions \cite{CarmichaelBook1}. In Eq. \eqref{Qfunc}, $\rho_{cv, {\rm ss}}$ is the reduced cavity density matrix $\rho_{cv, {\rm ss}}=\lim_{t \to \infty} [\braket{+|\rho(t)|+} + \braket{-|\rho(t)|-}]$, where $\ket{+}$ and $\ket{-}$ are the upper and lower states of the two-level atom, respectively \cite{SpontaneousDressed}. The {\it quasi}-probability distribution in the Q representation for a cavity field in the coherent state $\ket{\alpha_c}=\ket{x_c+iy_c}$ with average photon occupation $\braket{n} \equiv \braket{a^{\dagger}a}=\left|\alpha_c\right|^2$ assumes the Gaussian form: $Q_c(x+iy)=(1/\pi)\exp\{-[(x-x_c)^2 + (y-y_c)^2]\}$ \cite{CarmichaelBook1}.
\begin{figure}
\begin{center}
\includegraphics[width=8.75cm]{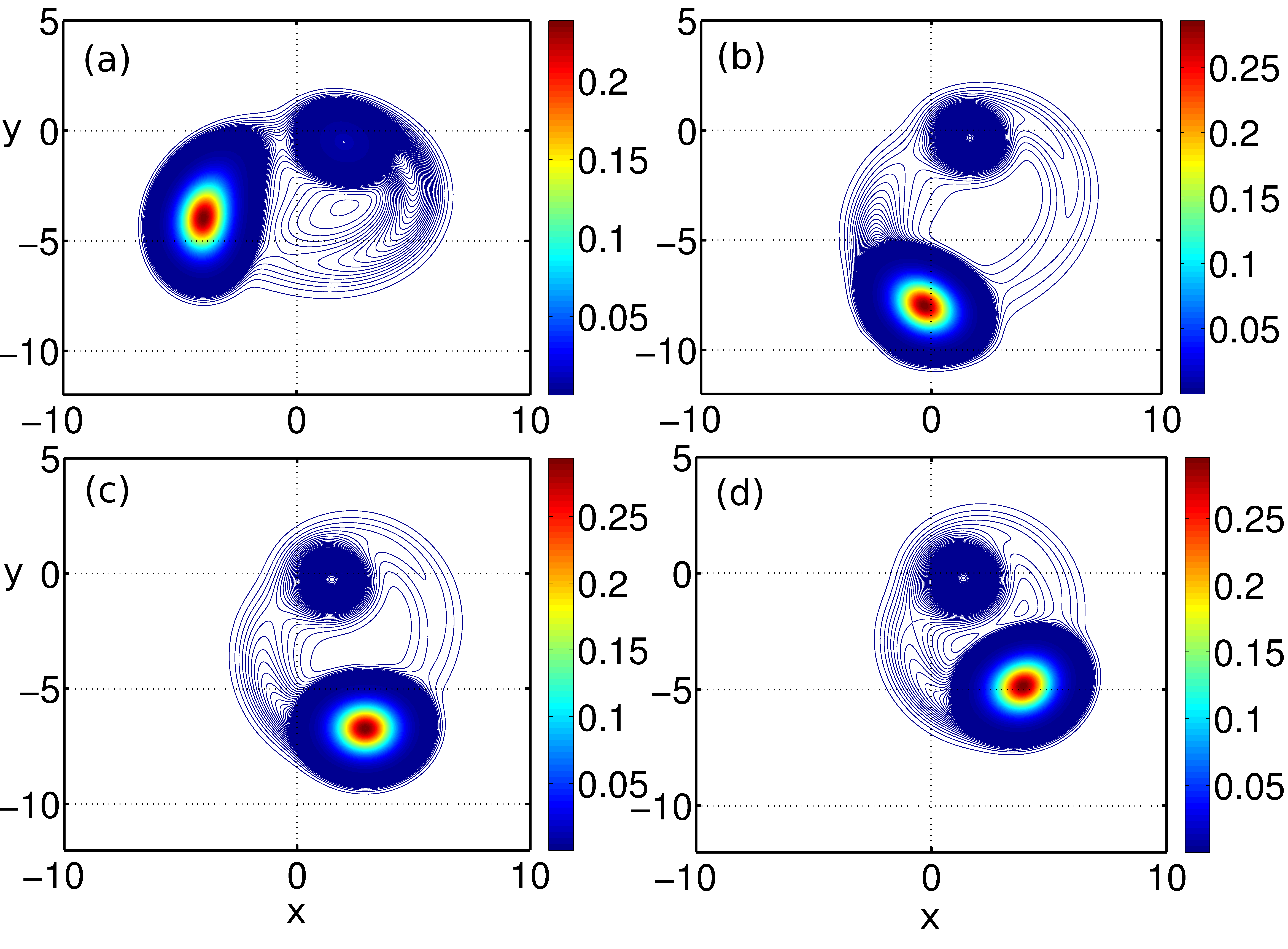}
\end{center}
\caption{{\bf Complex amplitude bistability shaping an excitation ring.} {\it Quasi}-probability function $Q(x+iy)$ of the intracavity field for varying drive-cavity detuning $\Delta\omega_c/\kappa$: $0.40, 0.96, 1.52, 2.08$ in {\bf (a)}-{\bf (d)} respectively. Parameters: $g/\kappa=16$, $\delta/g=-1.25$, $\gamma/(2\kappa)=0$ and $\varepsilon_d=g/2$.}
\label{fig:ring}
\end{figure}

Coupling the JC oscillator to a Markovian bath leads to the decay of coherence and degradation of quantum correlations, rendering ultimately the system amenable to a classical description \cite{Zurek}. Nevertheless, driving the cavity (or the qubit) with an external coherent field gives rise to non-trivial steady states with dynamics organized by quantum fluctuations in the strong coupling regime, investigated both theoretically and experimentally (see for example \cite{BishopRabi, SpontaneousDressed, CarmichaelBook2}). For $\delta=0$, ``photon blockade breaks down by means of a first-order dissipative quantum phase transition, except at a critical point in the space of drive amplitude and (drive-cavity) detuning, where a continuous transition is observed''\cite{CarmichaelPhotonBlockade}. Such a second-order transition is marked by spontaneous symmetry breaking in which the phases of the emergent states become nonlinear functions of the ratio $\varepsilon_d/g$ (with $\varepsilon_d \geq g/2$) \cite{CarmichaelBook2}. 

Interaction with a strong drive prevails over dressing of the JC system coupling (between the coupled qubit and cavity) which becomes a perturbation past a set threshold. While the states of phase bistability above threshold are nonstable in the mean-field, in the full quantum mechanical picture they become attractors with two isolated distributions in the $Q$ function for the intracavity field \cite{SpontaneousDressed, CarmichaelBook2}. In this work, we study the region of this critical point in the parameter space of the drive ($\Delta\omega_c/\kappa, \varepsilon_d/\kappa$) allowing a variation of the qubit-cavity detuning. We visit key results of the neoclassical theory of radiation alongside the r\^{o}le of quantum bistability and the associated scale parameters that connect us to the notion of a thermodynamic limit in a phase transition. 

Increasing the detuning between the resonant cavity mode and the laser driving field results in a rotation of an excitation ring in the phase portrait $(x, y)$ with varying intracavity photons, as we can see in Fig. \ref{fig:ring}, where the drive strength is set to its threshold value $\varepsilon_d=g/2$. For the smallest detuning, in Frame (a), the two states point to phase bistability attained for $\delta=0$ and $\Delta \omega_c=0$. The rotation of the most probable state in the half-plane $y<0$ for increasing $\Delta \omega_c>0$ corresponds to the intracavity photons following a Lorentzian curve with variable detuning, depending on the number of photons in the resonator [see Eq. (24) of \cite{CarmichaelPhotonBlockade} for cavity-qubit resonance]. 

The less probable state follows always its ``phase-bistable original counterpart'', a more probable brighter state, as a very low amplitude distribution in the $Q$ function \textemdash{an} evidence of quantum fluctuation switching between a resonant excitation with a Lorentzian spectrum, and a darker cavity state. The Maxwell-Bloch equations for $\gamma/(2\kappa) \to 0$ {\it do not predict any bistability} for all the drive parameters of Fig. \ref{fig:ring}, but only a bright state with $60, 42, 28, 19$ photons in (a)-(d) respectively, in quantitative and qualitative contrast to the trend exhibited by the quantum dynamics (with $\braket{n}_{\rm ss} \approx 30, 63, 53, 38$ respectively). 
\begin{figure}[!hb]
\begin{center}
\includegraphics[width=8.75cm]{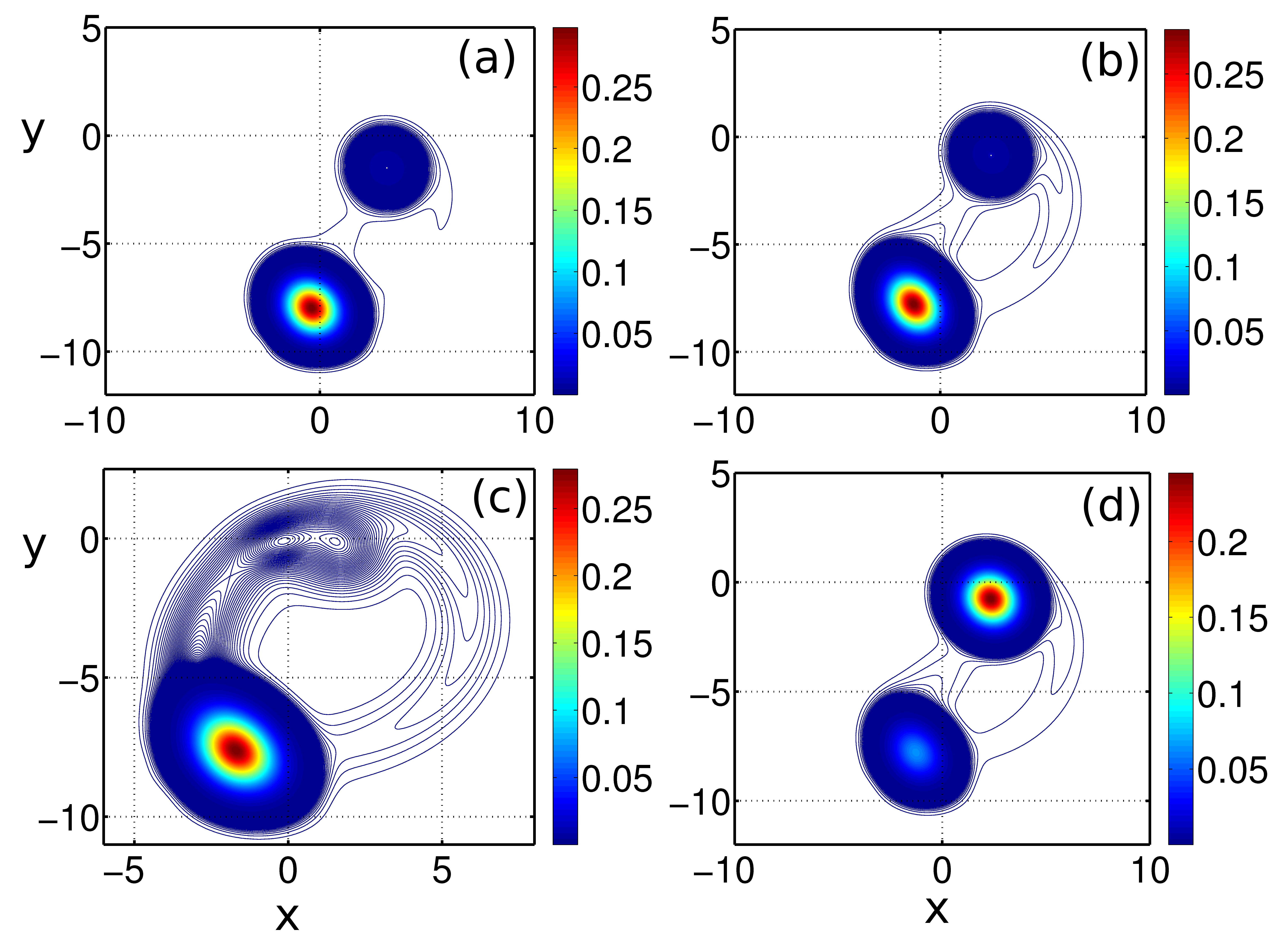}
\end{center}
\caption{{\bf Towards qubit-cavity resonance.} {\it Quasi}-probability function $Q(x+iy)$ of the intracavity field for varying cavity-qubit detuning $\delta/g$: $-10, -5, 0, +5$ in {\bf (a)}-{\bf (d)} respectively. Parameters: $\Delta\omega_c/\kappa=0.8$,  $g/\kappa=16$, $\gamma/(2\kappa)=0$ and $\varepsilon_d=g/2$.}
\label{fig:appres}
\end{figure}
As we can observe in Fig. \ref{fig:appres}, there is a sharp drop in the cavity photon number as we move from $\delta<0$ to $\delta>0$ since the probabilities of occupying the two neoclassical states are reversed, with the low-photon state (closer to the center of co-ordinates) becoming dominant [see Frame (d)]. At the same time, the states of complex-amplitude quantum bistability remain centered at the same positions in the phase portrait for the same $|\delta|$. Moreover, at $\delta=0$ [Frame (c)] there appears a third state very close to the center of co-ordinates along the excitation path of the JC ladder. This very low amplitude state is a prediction of the {\it neoclassical} theory of radiation, satisfying the state equation \cite{CarmichaelPhotonBlockade}:
\begin{equation}\label{neocl}
\begin{aligned}
&\alpha=-i\varepsilon_d \left[\kappa-i\left(\Delta\omega_c - \frac{g^2}{\sqrt{\Delta \omega_c^2 + 4g^2|\alpha|^2}} \right) \right]^{-1}\\
& \approx -i\varepsilon_d \left[\kappa-i\left(\Delta\omega_c - \frac{g^2}{\Delta\omega_c} \right) \right]^{-1}, \,\, {\rm Re}(\alpha) \approx -\frac{\varepsilon_d \Delta \omega_c}{g^2}.
\end{aligned}
\end{equation} 
The approximation in the second line would also give the Rabi resonances at $\Delta \omega_c=\pm g$ in the linear regime, for a much weaker drive and a larger drive-cavity detuning. 

\begin{figure}
\includegraphics[width=0.49\textwidth]{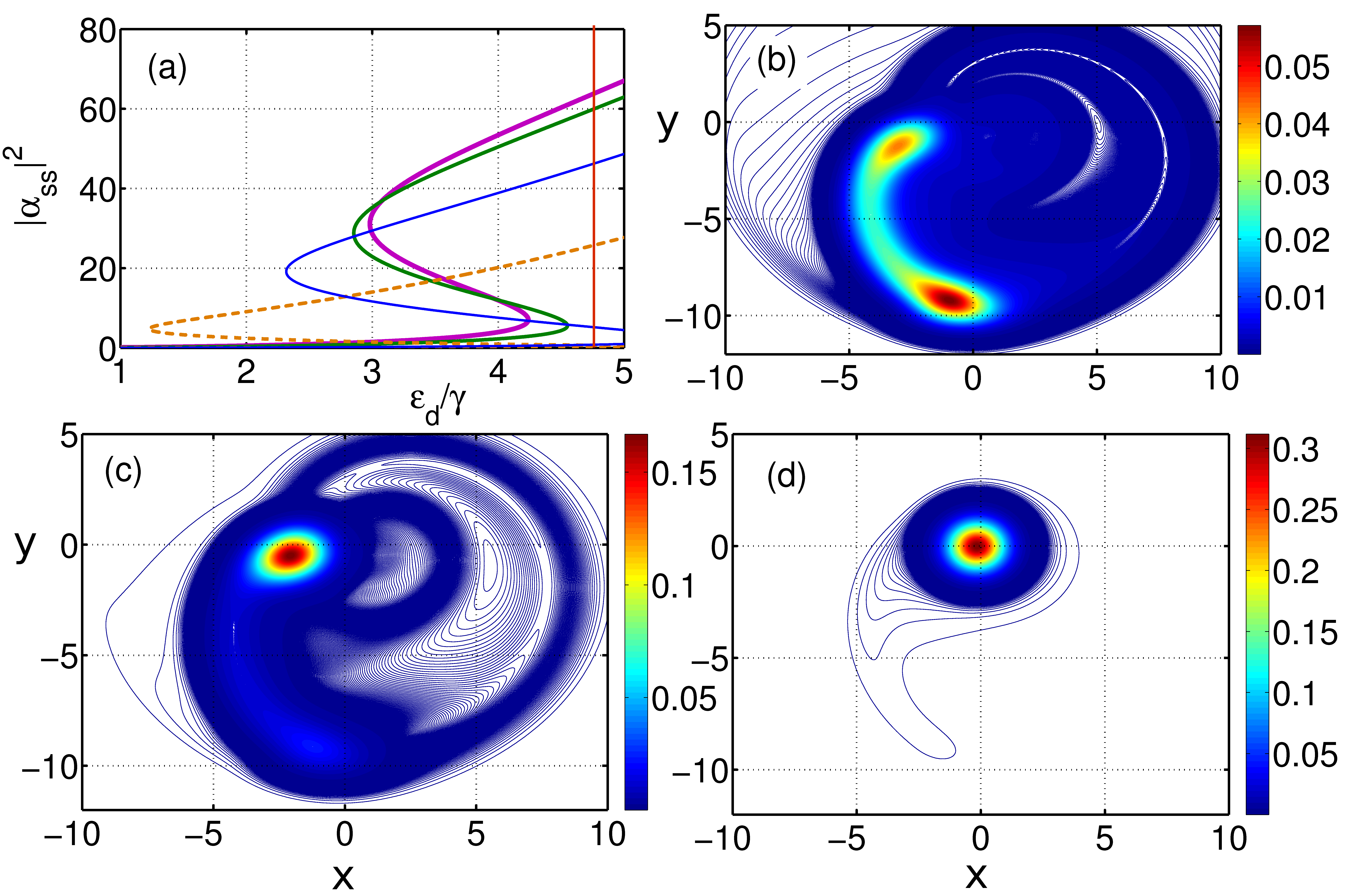}
\caption{{\bf Maxwell-Bloch and quantum bistability below threshold.} {\bf (a)} Semiclassical photon number $|\alpha_{\rm ss}|^2$ as a function of the normalized drive field $\varepsilon_d/\gamma$ for three different values of $\delta/g$: $-5.25$ (solid very thick line \textemdash{in} purple), $-4.5$ (solid thick line \textemdash{in} green), $-2.5$ (solid thin line \textemdash{in} blue) $-0.5$ (dashed line \textemdash{in} orange). {\bf (b)-(d)} {\it Quasi}-probability function $Q(x+iy)$ of the intracavity field for varying cavity-qubit detuning $\delta/g$: $-5.25, -4.5, -0.5$, respectively, and $\varepsilon_d/g=0.24$ [marked by the vertical line in (a) \textemdash{in} red]. Parameters: $\gamma/(2\kappa)=1$, $g/\gamma=20$ and $\Delta\omega_c/\kappa=2$.}
\label{fig:MFQuant}
\end{figure}

The two states in Frames (a, b, d) satisfy the mean-field state equation of the Kerr nonlinearity \cite{Bishop2010, CarmichaelPhotonBlockade}:
\begin{equation}\label{Kerrnonlin}
\alpha=-i\varepsilon_d \left\{\kappa-i\left[\Delta\omega_c + \frac{g^2}{\delta} \left(1 + \frac{4g^2}{\delta^2}|\alpha|^2\right)^{-1/2} \right] \right\}^{-1},
\end{equation}
one for $\delta<0$ (high-photon) and one for $\delta>0$ (low-photon state). We note, remarkably, that both states are present in the phase portrait {\it quasi}-distribution (quantum dynamics), even if the value of $\delta$ has a definite sign, while the variation of qubit-cavity detuning results only in the change of their relative weights. At the same time, the Maxwell-Bloch equations again do not predict any bistability for the corresponding drive parameters and vanishing spontaneous emission rate (compare also with Fig. \ref{fig:MFQuant}). The difference between the asymptotic behaviour of the Maxwell-Bloch state equation for bistability and the solution of the neoclassical equations has been discussed in detail for resonance ($\delta=\Delta\omega_c=0$) in \cite{SpontaneousDressed}, arising when considering the limit $\gamma/(2\kappa) \to 0$ {\it a priori} and {\it a posteriori} to forming the steady-state response.

For $\Delta \omega_c=0$, setting $\delta=\pm |\delta|$ in Eq. \ref{Kerrnonlin} yields two complex-conjugate neoclassical field amplitudes $i \alpha$. Taking now the limit $|\delta| \to 0$ recovers the two states of phase bistability, $\alpha=-i \varepsilon_d [\kappa \pm i g/(2|\alpha|)]^{-1}$, which is reflected by two symmetrically located peaks of equal height in the $Q$ function \cite{SpontaneousDressed}. In the opposite limit, when $|\delta| \gg g$ and $|\alpha|^2 \ll n_{\rm nc, \,Kerr}$, where $n_{\rm nc, \,Kerr}=[\delta/(2g)]^2$, the resonances of the linear strongly dispersive regime are located at $\Delta \omega_c=\pm g^2/|\delta|$ \cite{Blais2004, Koch2007, DispersiveQED}. We note that the scaling of Eq. \eqref{Kerrnonlin} appears only for the lowest-order Kerr nonlinearity in the case of weak dispersion at resonance ($\delta=0$), after a Taylor expansion of the root in the first line of Eq. \eqref{neocl}, matching the neoclassical predictions for $(\sqrt{2}g|\alpha_{\rm ss}|)^2 \ll \Delta \omega_c^2$ \cite{AgrawalCarmichael, CarmichaelPhotonBlockade}. More importantly, Eq. \eqref{neocl} brings along the scale parameter $n_{\rm sc}=g^2/(4\kappa^2)$ associated with a strong-coupling limit \cite{CarmichaelPhotonBlockade}. The low-photon state of Fig. \ref{fig:appres}(c) is linked to the states predicted by Eq. \ref{Kerrnonlin}, and is marked by strong quantum fluctuations, despite having $|\alpha|^2 \ll n_{\rm sc}$. 

The disparity between the mean-field predictions and the ME results persists even when we consider spontaneous emission into the modes of the vacuum field with a rate such that $\gamma/(2\kappa)=1$. At the driving field strength about half of its threshold value, the semiclassical bistability region is crossed when $\delta/g \to 0^{-}$, as we can see in Fig. \ref{fig:MFQuant} [frame (a)]. However, quantum bimodality gradually disappears in the associated $Q$ function [Frames (b)-(d)]. The bright state \textemdash{} the one furthest away from the center of co-ordinates \textemdash{} recedes along the excitation path of the JC ladder and the cavity photon number decreases constantly with diminishing qubit-cavity detuning. Due to the presence of appreciable spontaneous emission, mixing of the states participating in the quantum dynamics takes us to a larger area in the phase portrait in Frames (b) and (c), receding together with bistability. We note that here the dim state \textemdash{}the one closer to the center of co-ordinates \textemdash{} is now situated on the half-plane $x<0$, in contrast to the bistability with $\gamma=0$ (Figs. \ref{fig:ring} and \ref{fig:appres}). As $\delta/g \to 0^{-}$, the dim state comes to better agreement with the prediction of the Maxwell-Bloch bistability steady-state equation \cite{WallsBook}:
\begin{equation}\label{MBeq}
\alpha=-\frac{i\varepsilon_d}{\tilde{\kappa}} \left[1 + \frac{2g^2/(\tilde{\kappa} \tilde{\gamma})}{1 +  8g^2|\alpha|^2/|\tilde{\gamma}|^2} \right]^{-1},
\end{equation} 
where $\tilde{\kappa}=\kappa - i \Delta \omega_c$ and  $\tilde{\gamma}=\gamma - 2i\Delta \omega_q$, lying on the lower branch of the bistability curve. 

\begin{figure}
\begin{center}
\includegraphics[width=0.47\textwidth]{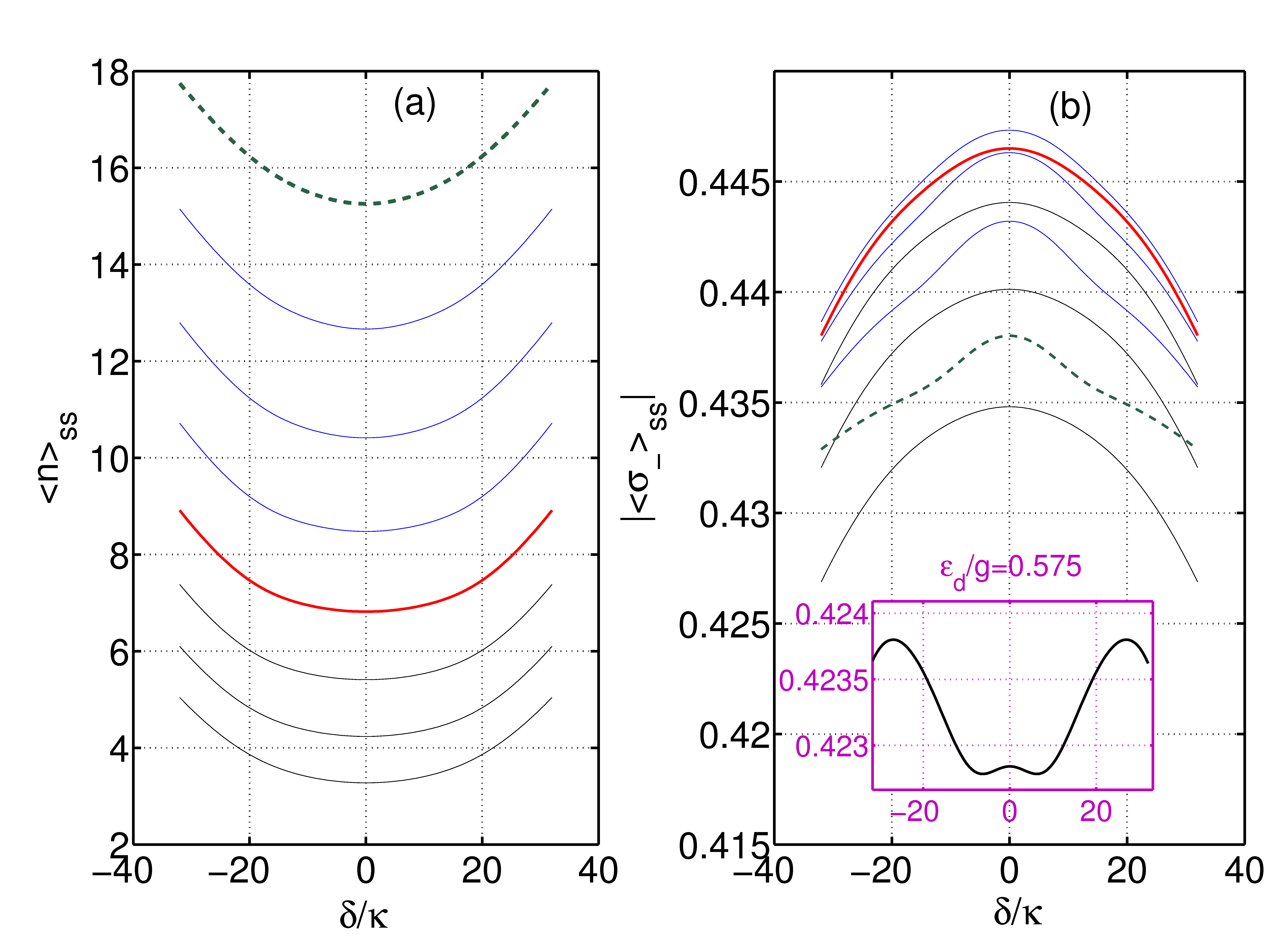}
\end{center}
\caption{{\bf Coupled, variably detuned cavity and qubit.} Average cavity photons
{\bf (a)} and modulus of the qubit coherence {\bf (b)} as a function of
$\delta/\kappa$ for eight equidistant values of the drive strength
$\varepsilon_d/g \in (0.45-0.55]$ [the inset in (b) is plotted for
$\varepsilon_d/g=0.575$]. The black (blue) curves correspond to drive
strengths below (above) threshold. The bold curve (in red) in both plots
indicates the threshold value $\varepsilon_d/g=0.5$ while the broken line
indicates the end of the range. Parameters: $\Delta \omega_c=0$,
$\gamma/(2\kappa)=0$ and $g/\kappa=16$.}
\label{fig:Vareps}
\end{figure}

Let us now examine side by side the two constituents of the JC oscillator when $\gamma/(2\kappa)=0$. In Fig. \ref{fig:Vareps} we present the average photon number $\braket{n}_{\rm ss}=\braket{a^{\dagger}a}_{\rm ss}$ extracted from the steady-state of the cavity density matrix, together with the modulus of the qubit coherence $\left|\braket{\sigma_{-}}_{\rm ss}\right|$ [with $\sigma_{-}=(1/2)(\sigma_{x}-i \sigma_{y})$], proportional to the distance from the $z$-axis in the Bloch sphere, $\sqrt{X^2+Y^2}$, both plotted as a function of the qubit-cavity detuning for varying drive strength. 
The photon curves are symmetric with respect to the sign of $\delta$, as Eq. \eqref{Kerrnonlin} suggests for $\Delta\omega_c=0$, while their minima correspond to the neoclassical asymptotic formula in the steady state \cite{CarmichaelPhotonBlockade}:
\begin{equation}\label{reseq}
\left|\alpha\right|^2=\frac{\varepsilon_d^2}{\kappa^2 + [g/(2\left|\alpha\right|)]^2},
\end{equation}
an approximation whose quality depends on the value of the drive strength relative to the threshold. The presence of cavity-drive detuning (for $\delta=0$) is linked to the split Lorentzian \cite{CarmichaelPhotonBlockade}, with the second term in the denominator of Eq. \eqref{reseq} becoming $[\Delta\omega_c \mp g/(2\left|\alpha_{\rm ss}\right|)]^2$. The nonlinearity is canceled at $\Delta \omega_c=\pm g\kappa/(2 \varepsilon_d)$, where $\left|\braket{\sigma_{-}}_{\rm ss}\right| \to 1/2$ when $\varepsilon_d/\kappa \gg 1$ [see Eq. (17) of \cite{CarmichaelPhotonBlockade} with $\left|\alpha_{\rm ss}\right|=\varepsilon_d/\kappa$, the empty-cavity steady-state excitation]. The limit value of $1/2$ is a prediction of the neoclassical theory above threshold at resonance ($\delta=\Delta\omega_c=0$), meaning that the qubit inversion is zero for every value of the {\it quasi}-energy in the continuous JC spectrum \cite{SpontaneousDressed}. As we can observe in Fig. \ref{fig:Vareps}, in the quantum picture the photon number increases at a steeper rate above threshold, while the qubit vector lies close the equatorial plane of the Bloch sphere, with the corresponding curve reversing trend and developing an inflection point, precursor of a curvature change [see inset of frame (b)].  

The Maxwell-Bloch bistability state equation \eqref{MBeq} indicates an effective co-operativity [$g^2/(\tilde{\kappa} \tilde{\gamma})$] as well as a scale parameter $n_{{\rm MB}}=\left|\tilde{\gamma}\right|^2/(8g^2)$. For $n_{\rm \,MB} \to \infty$, a weak-coupling limit for the open system, the displayed nonlinearity becomes essentially classical and quantum fluctuations play only the r\^{o}le of a negligible perturbation \cite{CarmichaelPhotonBlockade, CarmichaelBook2}. Alike is the scaling behaviour suggested by Eq. \eqref{Kerrnonlin}, since $n_{\rm nc, \,Kerr} \to \infty$ for $g \to 0$. For $\Delta \omega_q=0$, the scale parameter $n_{\rm \,MB}=\gamma^2/(8g^2)$ corresponds to the absorptive bistability discussed in \cite{Bonifacio} and \cite{singleatom} for many atoms and a single atom, respectively, coupled to a resonant cavity mode. 

We have already noted in Fig. \ref{fig:MFQuant} that the presence of a very small $\gamma^2/(8g^2)$ [$\ll \delta^2/(4g^2)$] suffices to alter the picture of bistability we have met in the preceding figures, both in terms of the peak locations and quantum fluctuations. For $\gamma/g \to 0$, the competition between dynamics of different scaling is marked by the pronounced coexistence of the states of Maxwell-Bloch bistability and the {\it structurally unstable} neoclassical states dependent upon the conservation of the qubit state vector in the Bloch sphere representation \cite{SpontaneousDressed, CarmichaelBook2}. A similar phenomenon is observed in a collection of classical phase oscillators non-locally coupled, where varying their coupling strength leads to the competition between structurally stable and turbulent complex states (called {\it chimeras}), an effect that can be viewed as a competition between coherence and decoherence \cite{BreatherChim, Panaggio, 1DChim}.

In conclusion, varying the detuning between the JC oscillator constituents and the drive allows us to extract information on the departure from the semiclassical theory, bringing together the dispersive and resonant behaviour around the critical point of a second-order phase transition. The effective JC nonlinearity, which depends upon the drive-induced dressing and dissipation, is dominated by quantum fluctuations at zero temperature through a competition of scaling dynamics governed by different parameters (and asymptotic dynamics when limits of these parameters are considered). Such a competition favours either two weak-coupling limits or one strong-coupling limit, bringing about a dissipative quantum phase transition with fundamental differences from its counterparts in many-body quantum optics, such as the laser and the superradiant phase transition. 

{\it Acknowledgements:} Simulations were performed using the Quantum Optics Toolbox in Matlab. The author thanks H. J. Carmichael for his guidance. The work was supported by the Engineering and Physical Sciences Research Council (EPSRC), UK.

\end{document}